\documentclass[pdftex,twocolumn,epjc3]{svjour3}          

\RequirePackage[T1]{fontenc}

\smartqed  

\RequirePackage{mathptmx}      
\RequirePackage{flushend}
\RequirePackage[numbers,sort&compress]{natbib}

\usepackage{graphicx,subfig}
\usepackage{float}
\usepackage{bm}
\usepackage{amsmath}
\usepackage{amsfonts,amssymb}
\usepackage{color}
\definecolor{v}{rgb}{0.6, 0.2, 0.8} 
\usepackage{soul} 
\usepackage{enumerate}
\usepackage{float}
\usepackage{subfig}
\usepackage{tabularx}
\usepackage{multicol}        
\usepackage{bm}	
\usepackage[draft]{hyperref}

\journalname{Eur. Phys. J. C}

\begin{document}

\title{Cosmological test on viscous bulk models using Hubble Parameter measurements and type Ia Supernovae data}


\author{A. Hern\'andez-Almada\thanksref{e1,addr1}
}

\thankstext{e1}{e-mail: ahalmada@uaq.mx}
\institute{Facultad de Ingenier\'ia, Universidad Aut\'onoma de Quer\'etaro, Centro Universitario Cerro de las Campanas, 76010, Santiago de Quer\'etaro, M\'exico \label{addr1}
}

\date{Received: date / Accepted: date}

\maketitle

\begin{abstract}
From a phenomenological point of view, we analyze the dynamics of the Universe at late times by introducing a polynomial and hyperbolic bulk viscosity into the Einstein field equations respectively. We constrain their free parameters using the observational Hubble parameter data and the Type Ia Supernovae dataset to reconstruct the deceleration $q$ and the jerk $j$ parameters within the redshift region $0<z<2.5$. At current epochs, we obtain 
$q_0 = -0.680^{+0.085}_{-0.102}$ and 
$j_0 = 2.782^{+1.198}_{-0.741}$ for the polynomial model and 
$q_0 = -0.539^{+0.040}_{-0.038}$ ($-0.594^{+0.056}_{-0.056}$) and 
$j_0 = 0.297^{+0.051}_{-0.050}$ ($1.124^{+0.196}_{-0.178}$) for the tanh (cosh) model. Furthermore, we explore the statefinder diagnostic that gives us evident differences with respect to the concordance model (LCDM). According to our results this kind of models is not supported by the data over LCDM.
\end{abstract}



\section{Introduction}
Currently the Universe is into an accelerated expansion phase supported by several cosmological observations coming from type Ia supernovae (SNIa)  \cite{Riess:1998,Perlmutter:1999}, the large-scale structure (LSS) \cite{Abbott:2017wau}, cosmic microwave background radiation (CMB) \cite{Planck:2015XIII,Planck:2015XIV}, baryon acoustic oscillations (BAO) \cite{Alam:2017}. Together with the observations coming from spiral galaxies \cite{Persic:1996, Salucci:2001} and galaxy clusters \cite{Frenk:1996}, the Universe contains more matter than the observed one known as dark matter (DM), and is responsible for the structure formation. The simplest cosmological model, called $\Lambda$-Cold Dark Matter (LCDM), describes these phenomena as two components that constitute the dark sector and is estimated to be about $95\%$ of the Universe. The accelerated expansion is well described by a cosmological constant (CC) with an equation of state (EoS) $p =-\rho$, and the structure formation is due by dust matter ($p=0$). Despite its good agreement to cosmological observations, LCDM presents several problems at galactic scales and open questions about the CC origin. Therefore, alternative models have been emerging to solve such LCDM inconsistencies. 
To explain the DM, we have axions \cite{Preskill:1983,Duffy:2009} (and ref. therein), ultralight scalar particles\cite{Matos:2000,Urena:2000,Matos:2001pz,Rodriguez-Meza:2012,Mario,RodriguezMeza/CervantesCota:2004}, supersymmetry particles \cite{Catena:2014}, and among others.  However, the cosmic measurements are not able to determine if the dark sector is constituted by two dark components due to the gravity theories only estimate the total energy-momentum tensor. This is known as {\it the degeneracy problem} \cite{HuEisenstein:1999, Kunz}. 

Motivated by this problem, a plenty of models proposes to explain the dark sector as a unique component or fluid that behaves as DM at high redshift and as DE at low redshift to model the current acceleration of the Universe solving the degeneracy problem.  Between them we have the (Generalized) Chaplygin gas \cite{Chaplygin,Kamenshchik:2001,Bilic:2001,Fabris:2001, LU2009404} with EoS ($p=A/\rho^n$) $p =A/\rho$ where $A$ and $n$ are constants, logotropic dark fluid \cite{Chavanis:2016} with $p = A \log(\rho/\rho_p)$ where $\rho_p$ is the Planck constant. More recently has been appearing models that generalize the perfect fluids EoS as $p = -\rho + \rho \,{\rm sinc}\,(\rho_0/\rho)$ where $\rho_0$ is the energy density at current epochs \cite{Hova2017, Almada:2018}. It is interesting to see that these models propose alternatives to the CC EoS. 
On the other hands, an interesting mechanism of unifying DM and DE supposes a Universe filled with a viscous fluid instead of a perfect fluid \cite{Zimdahl:1996, Coley:1996}. In this framework, the accelerated expansion of the Universe is due to viscous fluid pressure instead of a CC. Thus this kind of Unified DM models also avoids the CC problems, such as the {\it the cosmological constant problem} and {\it coincidence problem} \cite{RevModPhys.61.1, Zlatev:1999}. For an interesting review of viscous fluids see \cite{Brevik:2017}.
Moreover, by taking into account viscous fluids, it is possible to avoid singularities at the future, called Big Rips \cite{Caldwell:2003, Brevik:2011}, that appears when the DE models are in the phantom region, {\it i.e.}, $p/\rho<-1$ \cite{Xin_He:2007}. 
In this context, there are mainly two approaches to address the bulk viscosity, the Eckart \cite{Eckart:1940} and the Israel-Stewart-Hiscock (ISH) \cite{Israel1979}, and both have advantages and disadvantages. For instance, besides the ISH approach solves the problem of the causality, {\it e.i.}, the perturbations are propagated with a finite speed, it is more complex than the Eckart's one that only are known some analytical solutions when the viscosity is assumed in the form $\xi = \xi_0 \rho^{s}$; In particular, for solutions when $s=1/2$ see for instance \cite{Chimento_1997,MCruz:2017, NCruz:2018, NCruz:2018arx}. An inconvenient of this form is that $\xi$ diverges at high densities or early epochs of the Universe.
Regarding the Eckart's formalism, it is the simplest one but is a non-causal theory where the perturbations in the viscous fluid are moved at infinite speed; however, there are proposals that solve this problem by including correction terms of $\mathcal{O}(1/c^{2})$ in the theory (see \cite{Disconzi:2015} for more details). Only some polynomial models of the bulk viscosity have been studied widely \cite{Murphy:1973, Belinskii:1975, Brevik2005, Xin-He:2009, Normann:2016, Elizalde:2017, Normann:2017} as function of the energy density  or redshift. The cosmological model with a constant bulk viscosity coefficient has been studied for instance in \cite{Brevik2005, Normann:2017}, and it could have problems at the early epochs of the Universe which it is into a turbulent state (see \cite{Normann:2017} for an interesting discussion about this point). Hence, we motivate this work to explore more complex functions of the bulk viscosity in the Eckart approach. 

In this work, we aim to revisit three cosmological viscous fluid models and constrain their free parameters by performing a Bayesian Markov Chain Monte Carlo (MCMC) analysis with the latest cosmological data of the Hubble parameter (OHD) and SNIa distances at the background level. We use the OHD and SNIa measurements collected by \cite{Magana:2018} and \cite{Scolnic:2017caz} respectively. The first viscous model consists of a Universe with a polynomial bulk viscosity as a function of the redshift proposed by \cite{Xin-He:2009}. This model is built as a generalized form of the constant bulk viscosity studied in \cite{Brevik2005}.
The second model contains a more complex form for the bulk viscosity that involves the $\tanh$ function that depends on the Hubble parameter $E$ and was proposed by \cite{FOLOMEEV200875}. Finally, as an alternative to tanh form, we explore the Universe dynamics by proposing a $\cosh$ form for the bulk viscosity coefficient. The last one is motivated by the fact that cosh function has been used widely as scalar potentials to study the dark sector. Then in our case, we will use to model the bulk viscosity. In these models, the bulk viscosity terms are introduced into Einstein field equations as an effective pressure and following the Eckart' approach \cite{Eckart:1940}.

This paper is structured as follows. In Sec. \ref{sec:Model}, we present the generals on the viscous fluid models under study. Section \ref{sec:Data} describes the OHD and SNIa sample to constrain the viscous fluid models and we also explain the configuration for the Bayesian statistical analysis. In Sec. \ref{sec:Results} we give a discussion. Finally, in Sec. \ref{sec:Summary} we present the conclusions of our results.

\section{Fluid Viscous model} \label{sec:Model}

We study the dynamics of the Universe considering a flat Friedmann-Roberson-Walker (FRW) metric, {\it i.e.} $ds^2 = -dt^2 + a(t)(dr^2 + r^2d\Omega^2)$,
where $a$ is the scale factor and $t$ is the cosmic time containing a viscous fluid. Thus we introduce the bulk viscosity $\Pi$ component as a pressure term into the energy-momentum tensor,
\begin{equation}
    T_{\mu\nu} = \rho U_\mu U_\nu + (p+\Pi) (g_{\mu\nu} + U_\mu U_\nu)\,.
\end{equation}
where $U^\mu=(1,0,0,0)$ in the co-moving coordinates. From the Einstein field equations, the Friedmann equations are
\begin{eqnarray}
    H^2 &=& \frac{\kappa^2}{3}\rho \,,         \label{eq:H}       \\
    \dot{H} + H^2 &=& -\frac{\kappa^2}{6}(\rho + 3\tilde{p}) \label{eq:adotdot} \,,
\end{eqnarray}
where $\kappa^2=8\pi G$, $H= \dot{a}/a$ is the Hubble parameter, $\tilde{p} = p + \Pi$ is the effective pressure and $\Pi = - 3 \xi H$, with $\xi$ the bulk viscosity coefficient. The definition of $\Pi$ is motivated by fluid mechanics which the viscosity phenemenon is related to the velocity, {\it e.i.}, $\dot{a}$ \cite{Xin-He:2009,Brevik:2017}.   As a first approximation, we have only considered one fluid in the model, however, by adding more components to the model such as radiation and dark energy and  assuming $\xi = \xi(H)$, the assigning the bulk viscosity to any fluid produces degeneracy, {\it i.e.}, it is not possible to identify which cosmic component is producing the viscosity at the background level \cite{Velten:2013}. On the other hand, the dust particles may produce viscosity through their decay to relativistic particles at low redshift ($z<1$) \cite{Turner:1984, Wilson:2007, Mathews:2008}. Such final particles must not be energetic photons because they would be detected easily \cite{Caldwell:2002}, instead they could be a kind of sterile neutrino or some other weakly interacting  candidate (for an interesting summary about such candidates, see \cite{Mathews:2008}). Hence we will consider that our cosmic component behaves as a pressureless dust-like matter ($p=0$).
Then the continuity equation is given by
\begin{equation} \label{eq:continuity2}
    \dot{\rho} + 3H\rho = 9\xi H^2 \,.
\end{equation}
We define the dimensionless Hubble parameter as
\begin{equation}\label{eq:H2def}
    E(t)^2 = \frac{H(t)^2}{H_0^2} = \frac{\rho}{\rho_{cr}} \,,
\end{equation}
where $\rho_{cr}$ is the critical density. From Eqs. (\ref{eq:H}), (\ref{eq:adotdot}), (\ref{eq:H2def}) and the relation $z = 1/a -1$ we obtain
\begin{equation}\label{eq:diffE}
    -2(1+z) \frac{dE}{dz} + 3E = 9\lambda\,,
\end{equation}
where we have defined $\lambda = \xi H_0/ \rho_{cr}$. It is interesting to notice that the Eq. (\ref{eq:diffE}) gives a correlation between deceleration parameter $q(z)$ and the dimensionless bulk viscosity $\lambda(z)$, then we can study this kind of models by proposing phenomenological functions to describe the parameter $q$ at late times. Hence, from Eq. (\ref{eq:diffE}) we can write the deceleration parameter as
\begin{equation}\label{eq:q}
    q(z)= \frac{1}{2} - \frac{1}{2E(z)}9\lambda(z) \,.
\end{equation}
Notice that it is interesting that this expression allows to propose a phenomenological behaviour of $\lambda(z)$ to model the accelerated dynamics of the Universe.

We also analyze the statefinder (SF) diagnostic that is useful to distinguish the behaviour of different cosmological models of LCDM model. The SF diagnostic is a $\{s,r\}$-plane defined by the geometric variables \cite{Sahni2003, Alam:2003} 
\begin{eqnarray}
r &=& j = \frac{\dddot{a}}{aH^3}\,, \\
s &=& \frac{r-1}{3(q-1/2)}\,,
\end{eqnarray}
where $r$ is the jerk parameter and $q$ is the deceleration one. In this phasespace, LCDM is a fixed point located at $(s,r)=(0,1)$, the trajectories in the region $r<1$ and $s>0$ corresponds to quintessence behaviour and trajectories in the region $r<1$ and $s<0$ presents a Chaplygin gas one. 

The jerk parameter can be expressed in terms of $q(z)$ and its first derivative with respect to $z$ as \cite{Mamon:2018}
\begin{equation}\label{eq:jerk}
    j(q) = q(2q+1) + (1+z)\frac{dq}{dz}\,.
\end{equation}
For the LCDM model, the jerk parameter is $j=1$ that it will be used to compare with our models.


\subsection{Polynomial function}
We consider the polynomial function of $\lambda$ 
given by \cite{Xin-He:2009}
\begin{equation} \label{eq:lambda1}
    9 \lambda(z) = \lambda_0 + \lambda_1 (1+z)^n \,,
\end{equation}
where $\lambda_0$, $\lambda_1$, and $n$ are free parameters to be determined by data sets. The simplest case, $\lambda = {\rm constant}$, was studied in \cite{Murphy:1973}.
Substituting Eq. (\ref{eq:lambda1}) in (\ref{eq:diffE}), we obtain the solution
\begin{equation} \label{eq:Epol}
    E(z) = \lambda_2 (1+z)^{3/2} - \frac{\lambda_1}{2n-3}(1+z)^n + \frac{\lambda_0}{3}\,,
\end{equation}
where
\begin{equation}
   \lambda_2 =  1  + \frac{\lambda_1}{2n-3} - \frac{\lambda_0}{3} \,.
\end{equation}
The deceleration parameter is given by
\begin{eqnarray}
q(z) &=& -1  \\ 
        & & + \frac{ \frac{3}{2} \lambda_2 (2n-3)(z+1)^{3/2}- \lambda_1 n(z+1)^{n}}{\frac{1}{3}\lambda_0(2n-3)+\lambda_2 (2 n-3)(z+1)^{3/2} - \lambda_1
   (z+1)^n }\,. \nonumber
\end{eqnarray}
Notice that for $n<0$, $q(z)\rightarrow 1/2$ when $z\rightarrow \infty$. The jerk parameter can be expressed as
\begin{eqnarray}
j(z) &=& 1 +  \frac{1}{E^{2}} \left( \frac{1}{2} \lambda_0^{2} + \lambda_0 \lambda_{1} (z+1)^n  + \frac{1}{2} \lambda_{1}^{2} (z+1)^{2 n} \right . \nonumber \\
     & &  - \frac{3}{2} \lambda_{0} E  - \frac{1}{2} \lambda_{1} (n  + 3 ) (z+1)^{n} E \nonumber \\
     & & \left.  + \frac{1}{2} \lambda_{0} ( z  + 1 ) \frac{dE}{dz} + \frac{1}{2} \lambda_{1} (z+1)^{n+1}\frac{dE}{dz} \right) \,,
\end{eqnarray}
where
\begin{equation}
    \frac{dE}{dz} = \frac{3}{2}\lambda_2 (z+1)^{1/2} - \frac{n}{2n-3}\lambda_1 (z+1)^{n-1} \,.
\end{equation}
Notice that the jerk parameter for the LCDM model is $j=1$.

\subsection{Hyperbolic function}

Motivated for describing the evolution of the Universe from recombination to late acceleration phase, the authors \cite{FOLOMEEV200875} proposed the hyperbolic behavior of $\lambda$ given by
\begin{equation} \label{eq:lambda2}
    9\lambda(z) = 3\tanh{\left(\frac{b}{E(z)^n} \right)}\,,
\end{equation}
where $b$ and $n$ are free parameters that we will determine by the cosmological data set.
The jerk parameter is expressed as
\begin{eqnarray}
j(z) &=&  1 + \frac{9}{4} b n E^{-n-1}  \nonumber \\
     & &- \left( \frac{9}{4} b n E^{- n - 2} + \frac{9}{4} E^{- 1}\right) \tanh (b E^{- n}  ) \nonumber \\ 
& &  - \left( \frac{9}{4} b n E^{-n-1} - \frac{9}{4} E^{-2}\right) \tanh^{2}{\left (b E^{- n} \right )} \nonumber \\ 
& & + \frac{9}{4} b n E^{- n - 2} \tanh^{3} (b E^{- n})\,,
\end{eqnarray}
where $E(z)$ will be obtained by solving differential equation numerically.

As alternative model to (\ref{eq:lambda2}), we study the phenomenological bulk viscosity model expressed as
\begin{equation} \label{eq:lambda3}
        9 \lambda(z) = \cosh{ \left(\frac{b}{E(z)^n} \right) } \,.
\end{equation}
The deceleration parameter is straightforward obtained from Eq. (\ref{eq:q}) and the jerk parameter can be expressed as
\begin{eqnarray}
j(z) &=&  1 + \frac{1}{8}E^{-2}  \nonumber \\
     & &  + \frac{3}{4} b n E^{- n - 1} \sinh{\left (b E^{- n} \right )}  - \frac{1}{8} b n E^{- n - 2} \sinh{\left (2 b E^{- n} \right )} \nonumber \\
     & &  - \frac{3}{4}E^{-1} \cosh{\left (b E^{- n} \right )} + \frac{1}{8}E^{-2} \cosh{\left (2 b E^{- n} \right )} \,.
\end{eqnarray}
Again, for the LCDM model we have $j=1$. The next section is devoted to describing the cosmological data used in this work.

\section{Cosmological data} \label{sec:Data}

In this section, we describe the observational datasets used to perform the confidence region of the free model parameters. We perform a Bayesian Chain Markov Monte Carlo analysis based on emcee module \cite{Emcee:2013} by setting $5000$ chains with $500$ steps. 
The nburn is stopped up to obtain a value of $1.1$ on each free parameters in the Gelman-Rubin criteria \cite{Gelman:1992}. We search the confidence region according to the priors presented in Table \ref{tab:priors} and using the Hubble parameter measurements and supernovae data. To compare the hyperbolic models with the data, we implement a second order Runge-Kutta procedure to solve the corresponding ODE with the functions given in Eqs. (\ref{eq:lambda2}) and (\ref{eq:lambda3}) respectively. For the polynomial viscous model, we use the exact solution of $E(z)$ given in Eq. (\ref{eq:Epol}).

\begin{table}
\caption{Priors considered for the polynomial (top panel) and hyperbolic (bottom panel) models. The priors for the polynomial model are based on \cite{Xin-He:2009}.}
\centering
\begin{tabular}{| cc  |}
\hline
Parameter    &  Prior                  \\
\multicolumn{2}{|c|}{Polynomial model} \\
\hline
$\lambda_0$  & Flat in $[0,2]$     \\ [0.7ex]
$\lambda_1$  & Flat in $[0,4]$     \\ [0.7ex]
$n$          & Flat in $[-5,0]$     \\ [0.7ex]
$h$          & Gauss$(0.7324,0.0174)$ \\ [0.7ex]
\hline
\multicolumn{2}{|c|}{tanh/cosh model}     \\
$b$          & Flat in $[0,3]$     \\ [0.7ex]
$n$          & Flat in $[0,5]$     \\ [0.7ex]
$h$          & Gauss$(0.7324,0.0174)$ \\ [0.7ex]

\hline
\end{tabular}
\label{tab:priors}
\end{table}

We perform a joint analysis by combining the OHD and SNIa data through the merit-of-function
\begin{equation}
\chi^2_{joint} = \chi^2_{OHD} + \chi^2_{SNIa}\,,
\end{equation}
where $\chi^2_{OHD}$ and $\chi^2_{SNIa}$ refer to the chi-square functions. The rest of the section is devoted to describing the observational data and the construction of each $\chi^2$ functions.

\subsection{Hubble Parameter Data}
The Universe is in an expansion rate that is measured through the Hubble parameter measurements (OHD). The OHD give are cosmological-model independent measurements of the Hubble parameter, $H(z)$, as a function of the redshift $z$ and the latest ones are obtained by the differential age (DA) tool \cite{Jimenez:2001gg} and BAO measurements. We consider the OHD compilation provided by \cite{Magana:2018} that consists of $51$ data points within the redshift region $[0,2.36]$. Thus we constrain the free parameters of the viscous models by building the chi-square as

\begin{equation}\label{eq:chi2_ohd}
\chi^2_{OHD} = \sum_{i}^{51} \left( \frac{H_{th}(z_i) - H_{obs}}{\sigma_{obs}^i} \right)^2 \,,
\end{equation}
where the $H_{th}(z_i)$ and $H_{obs}(z_i) \pm \sigma_{obs}^i$ are the theoretical and observational Hubble parameter at the redshift $z_i$ respectively.

\subsection{Type Ia Supernovae}

The Pantheon compilation given by \cite{Scolnic:2017caz} contains the observations of the luminosity modulus coming from $1048$ type Ia supernovae (SNIa) located in the region $0.01 < z < 2.3$. 
The free parameters of the model are obtained by minimizing the merit-of-function 
\begin{equation}
\chi^2_{SNIa} = (m_{th}-m_{obs}) \cdot {\rm Cov}^{-1} \cdot (m_{th}-m_{obs})^{T}\,,
\end{equation}
where $m_{th}-m_{obs}$ is the difference between the theoretical and observational bolometric apparent magnitude and ${\rm Cov}^{-1}$ is the inverse of the covariance matrix. The theoretical bolometric apparent magnitude is computed by
\begin{equation}
m_{th}(z) = \mathcal{M} + 5\, \log_{10}\left[ d_L(z)/10\,pc \right]\,.
\end{equation}
Here, $\mathcal{M}$ is a nuisance parameter and $d_L(z)$ is the dimensionless luminosity distance given by
\begin{equation}
d_L(z) =  (1+z)\,c \int_0^z \frac{dz'}{H(z')} \,,
\end{equation}
where $c$ is the speed of light.

\section{Discussions} \label{sec:Results}

\begin{figure}[H]
  \centering
  \includegraphics[width=.9\linewidth]{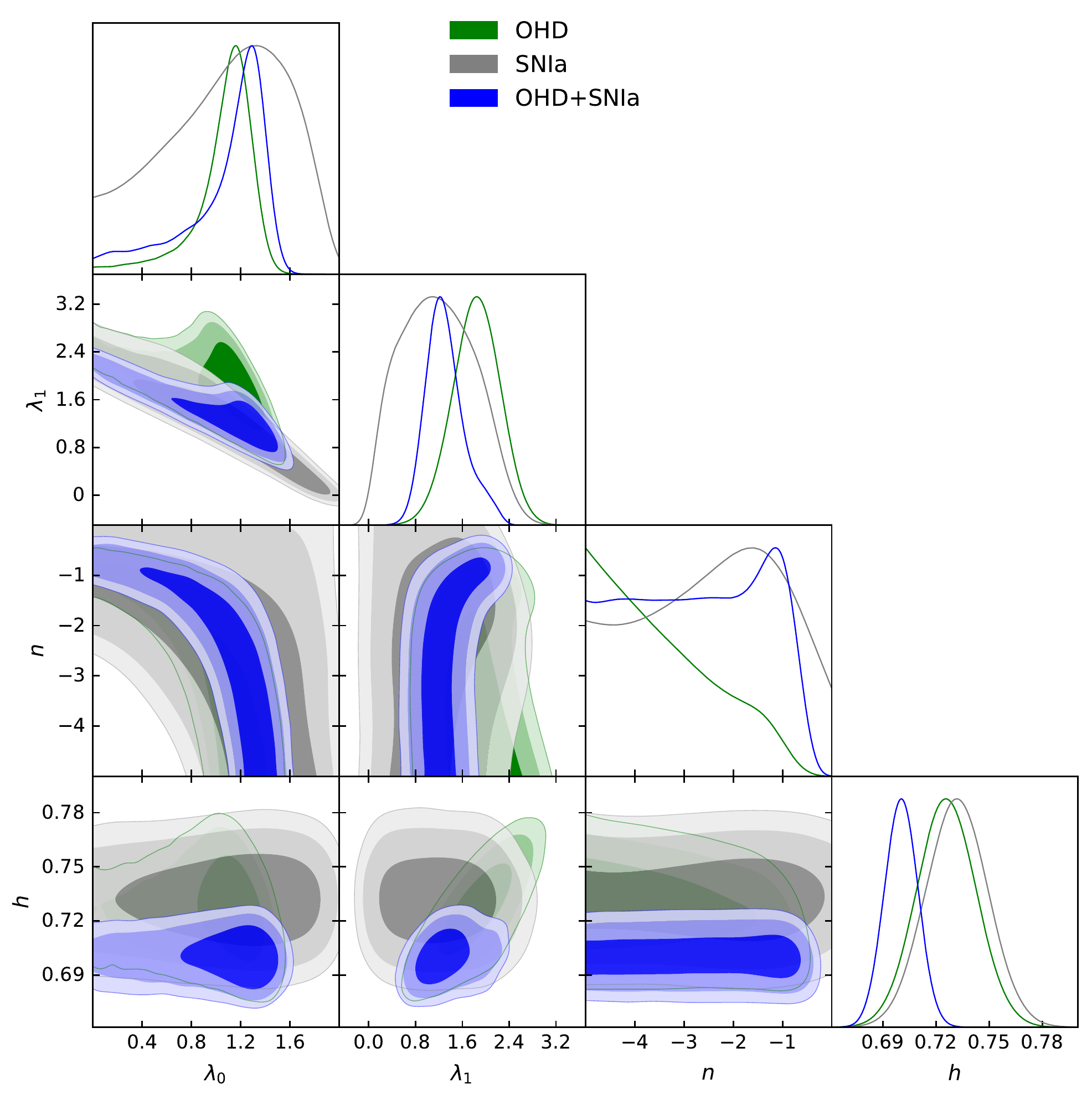}
\caption{$68$, $97$ and $99.7\%$ CL for free model parameters considering $9\lambda = \lambda_0 + \lambda_1 (1+z)^n$.}
\label{fig:contour1}
\end{figure}

\begin{figure}[H]
  \centering
  \includegraphics[width=.9\linewidth]{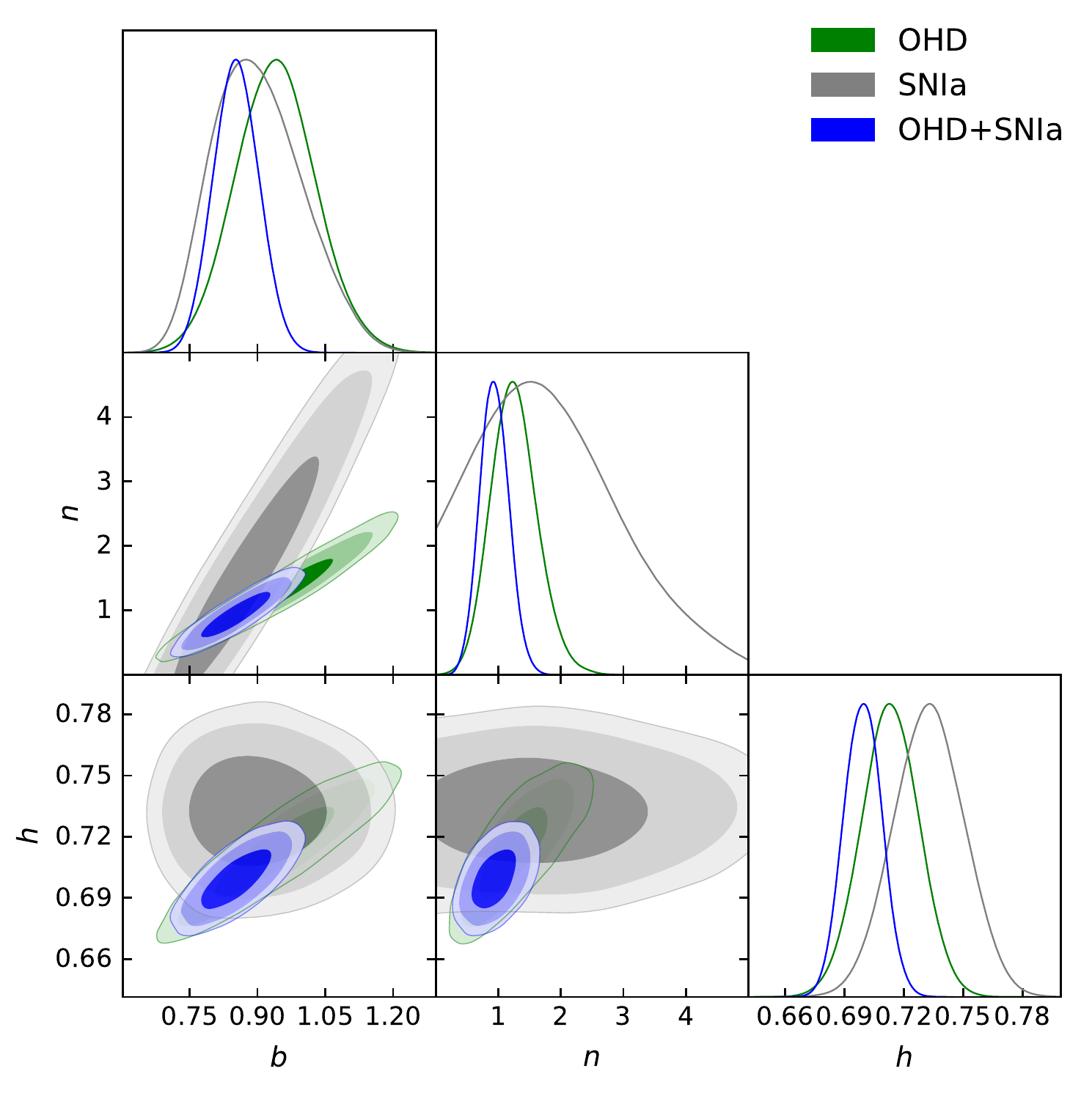}
\caption{$68$, $97$ and $99.7\%$ CL for free model parameters considering $9\lambda = 3\tanh{( bE(z)^{-n}) }$.}
\label{fig:contour2}
\end{figure}

\begin{figure}[H]
  \centering
  \includegraphics[width=.9\linewidth]{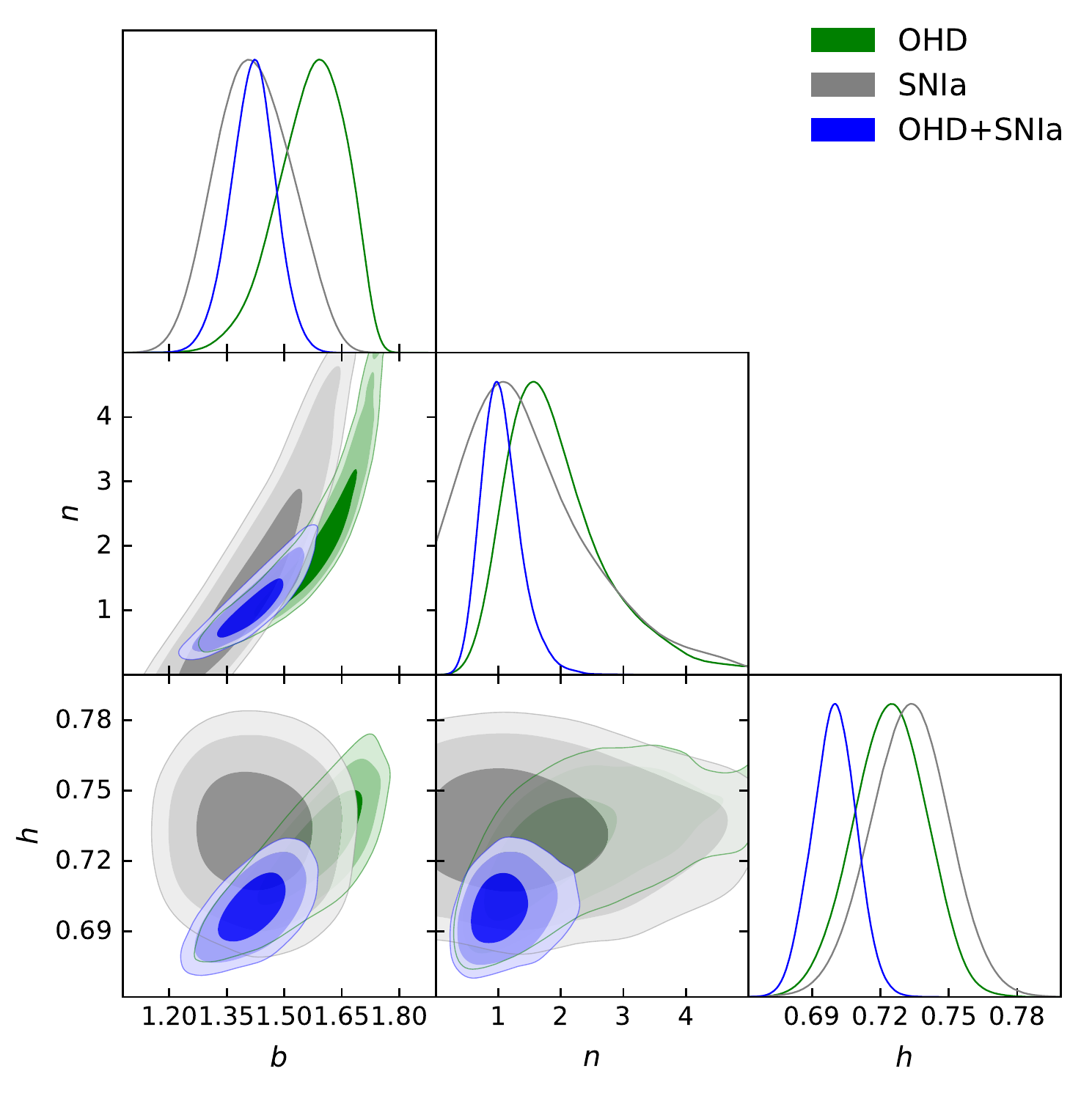}
\caption{$68$, $97$ and $99.7\%$ CL for the free model parameters considering $9\lambda = \cosh{( b E(z)^{-n}) }$.}
\label{fig:contour3}
\end{figure}

The best fit values of the free model parameters obtained by our OHD, SNIa and joint analysis are summarized in the Tab. \ref{tab:bf_model1} for the polynomial model and in the Tab. \ref{tab:bf_model23} for the hyperbolic models. The uncertainties showed correspond at $68\%$ ($1\sigma$) confidence Level (CL). Also, we present the $1\sigma$, $2\sigma$, $3\sigma$ CL 2D contours and 1D marginalized posterior distributions of the model parameters in Figs. \ref{fig:contour1}, \ref{fig:contour2}, and \ref{fig:contour3} for polynomial, tanh, and cosh models respectively using  OHD, SNIa and OHD+SNIa data.
From now, we use the best values obtained in the joint analysis for our discussions. For the polynomial model, \cite{Xin-He:2009} estimates best fitting yields around $\lambda_0\approx 0.64$, $\lambda_1\approx 1.83$ when the parameter $n$ is fixed at $n=-2$ in the fit (for more details see Table 1 of \cite{Xin-He:2009}). Our values are consistent within $1.6\sigma$ CL with those reported by \cite{Xin-He:2009}.
For tanh model, we obtain a yield value of the $b$ parameter consistent within $2\sigma$ CL and a deviation of about $4.5\sigma$ CL for the parameter $n$ with those reported by \cite{FOLOMEEV200875}.
 It is interesting to see that for all the models when $n=0$ we could recover the cosmological model with bulk viscosity coefficient.  We observe a deviation of a cosmological model with constant bulk viscosity  ($n=0$) from tanh and cosh model of about $4.5\sigma$ and $3.7\sigma$, respectively. Regarding to the polynomial model, we estimate a limit of $n<-0.54$ at $99\%$ CL. Also, notice for the latter model, it could be obtained a bulk viscosity coefficient to be constant by making $\lambda_1=0$, hence, we obtain a deviation of about $4.53\sigma$.

\begin{table*}
\caption{Best fitting parameters of the polynomial model.}
\centering
\begin{tabular}{| ccccccc |}
\hline
\multicolumn{7}{|c|}{Polynomial model} \\
Data     & $\chi^2$ &    $\lambda_0$            & $\lambda_1$               & $n$                        & $h$ & $\mathcal{M}$       \\
\hline
OHD      & $15.1$ & $1.112^{+0.154}_{-0.256}$ & $1.844^{+0.399}_{-0.408}$ & $-3.628^{+1.534}_{-0.990}$ & $0.726^{+0.017}_{-0.017}$ &  -  \\ [0.7ex]
SNIa     & $1027.9$ & $1.129^{+0.459}_{-0.618}$ & $1.159^{+0.702}_{-0.668}$ & $-2.351^{+1.374}_{-1.712}$ & $0.732^{+0.017}_{-0.017}$  & $5.741^{+0.053}_{-0.055}$ \\ [0.7ex]
OHD+SNIa & $1053.2$ & $1.183^{+0.177}_{-0.487}$ & $1.273^{+0.363}_{-0.281}$ & $-2.656^{+1.494}_{-1.596}$ & $0.700^{+0.009}_{-0.009}$  &  $5.634^{+0.023}_{-0.023}$ \\ [0.7ex]
\hline
\end{tabular}
\label{tab:bf_model1}
\end{table*}

\begin{table*}
\caption{Best fitting parameters of the tanh and cosh models.}
\centering
\begin{tabular}{| cccccc |}
\hline
 \multicolumn{6}{|c|}{Hyperbolic models} \\
  \hline
 \multicolumn{6}{|c|}{tanh model} \\
 
Data     &$\chi^2$  &    $b$                      & $n$                         & $h$                       & $\mathcal{M}$   \\
\hline
 OHD     & $28.8$    & $0.937^{+0.088}_{-0.087}$   & $1.230^{+0.376}_{-0.350}$   & $0.713^{+0.014}_{-0.015}$  & -   \\ [0.7ex]
SNIa     & $1026.3$    & $0.894^{+0.109}_{-0.094}$   & $1.727^{+1.271}_{-1.005}$   & $0.733^{+0.018}_{-0.018}$  &  $5.747^{+0.054}_{- 0.056}$ \\ [0.7ex]
OHD+SNIa & $1055.7$    & $0.853^{+0.050}_{-0.050}$   & $0.933^{+0.236}_{-0.222}$   & $0.699^{+0.009}_{-0.009}$  & $5.644^{+0.023}_{-0.023}$  \\ [0.7ex]
\hline
 \multicolumn{6}{|c|}{cosh model} \\
Data     &$\chi^2$  &    $b$                      & $n$                         & $h$        & $\mathcal{M}$                   \\
 \hline
 OHD     & $26.7$   & $1.580^{+0.084}_{-0.098}$   & $1.790^{+0.939}_{-0.605}$   & $0.724^{+0.016}_{-0.016}$ &  -   \\ [0.7ex]
SNIa     & $1041.5$ & $1.417^{+0.106}_{-0.096}$   & $1.348^{+1.225}_{-0.782}$   & $0.733^{+0.017}_{-0.017}$ &  $5.747^{+0.051}_{-0.053}$   \\ [0.7ex]
OHD+SNIa & $1054.7$ & $1.420^{+0.056}_{-0.059}$   & $1.014^{+0.339}_{-0.273}$   & $0.700^{+0.009}_{-0.010}$ &  $5.640^{+0.023}_{-0.024}$   \\ [0.7ex]
\hline
\end{tabular}
\label{tab:bf_model23}
\end{table*}

\begin{figure}[H]
  \centering
  \includegraphics[width=.9\linewidth]{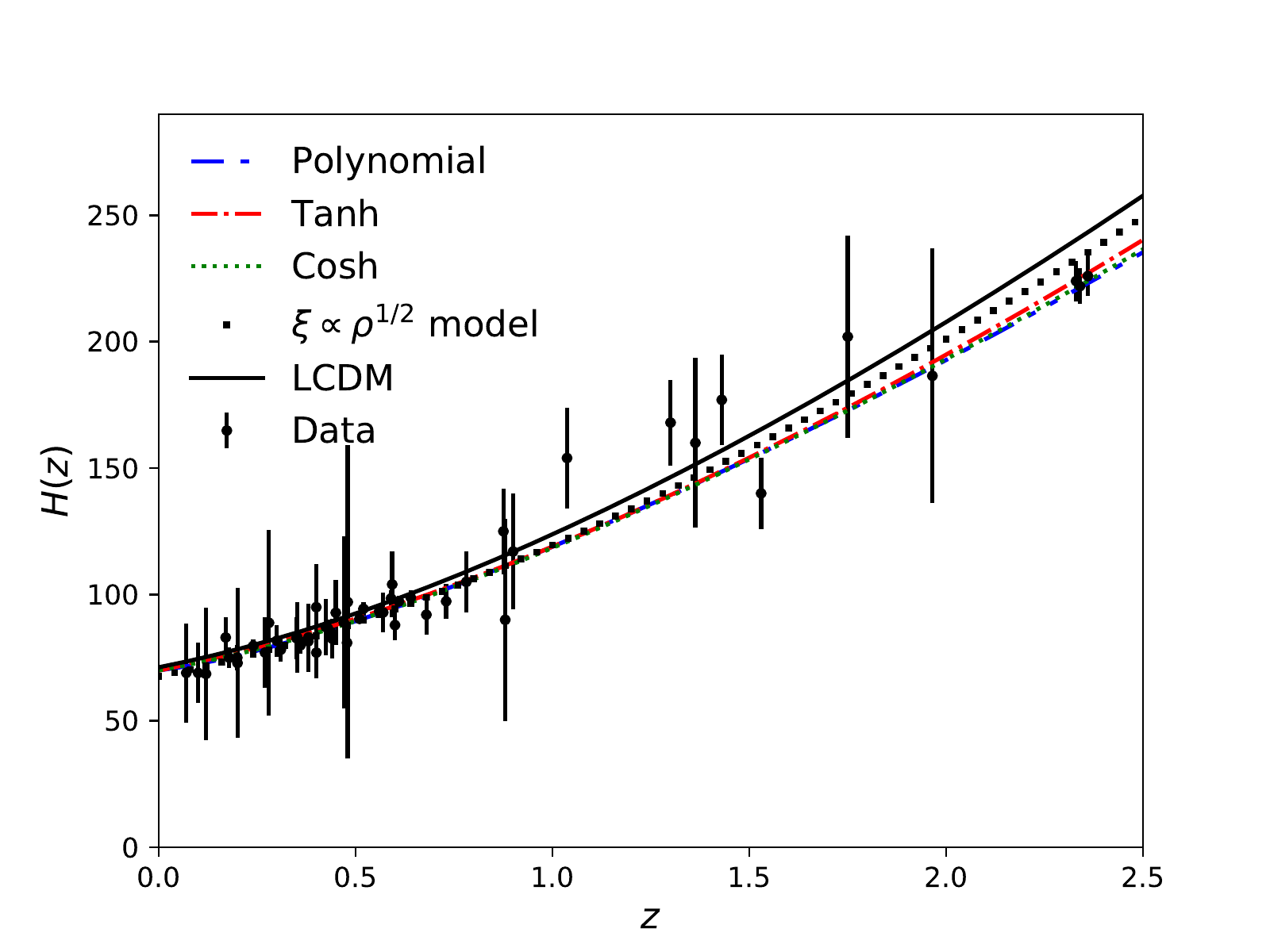}
\caption{Best fit curves using the results of the joint analysis for LCDM (black solid line), polynomial (blue double dash-dotted line), tanh (red dash-dotted line), and cosh (green dotted line) models. For LCDM, we use $\Omega_m = 0.289$, and $h=0.712$. The black squares correspond to the bulk viscosity of the form $\xi \propto \rho^{1/2}$ obtained by \cite{Normann:2016}. The black points with uncertainty bars correspond to OHD.}
\label{fig:Hz_fit}
\end{figure}

Figure \ref{fig:Hz_fit} displays the best curves of the models over $H(z)$ data. To compare with LCDM model we also performed a MCMC joint analysis to obtain its best fitting parameters. Taking into account $\Omega_r = 2.469\times 10^{-5} h^{-2}(1 + 0.2271 N_{eff})$ where $N_{eff}=3.04$, \cite{Komatsu:2011} the best fitting values are $\Omega_{m0} =  0.289^{+0.019}_{-0.018}$ and $h = 0.712^{+0.013}_{-0.013}$ with a $\chi^2 = 1043.7$. Also we compare our results with one obtained by \cite{Normann:2016} when the bulk  viscosity is considered in the form $\xi \propto \sqrt{\rho}$.

\begin{figure}[t]
  \centering
  \includegraphics[width=.9\linewidth]{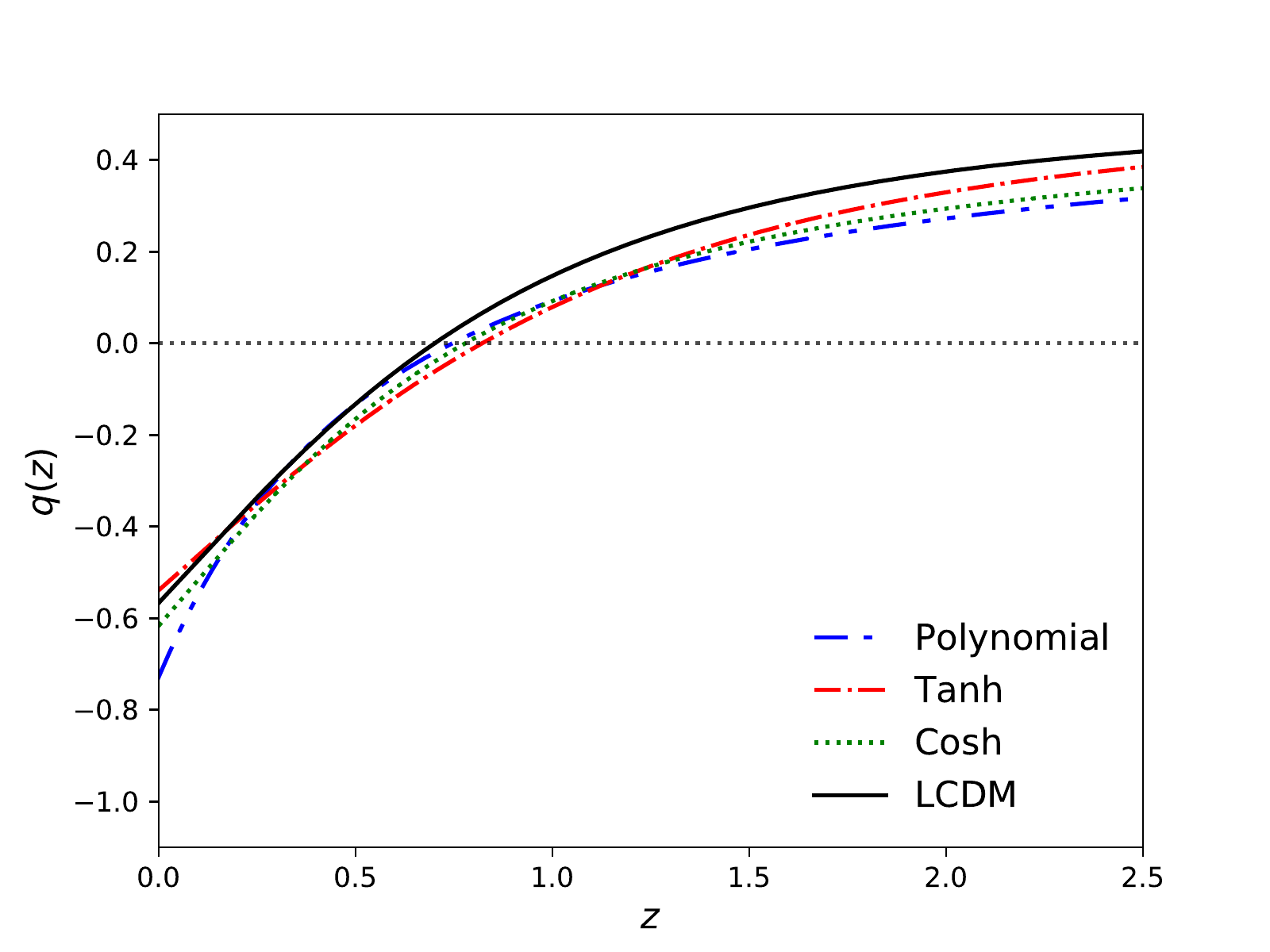}\\
  \includegraphics[width=.9\linewidth]{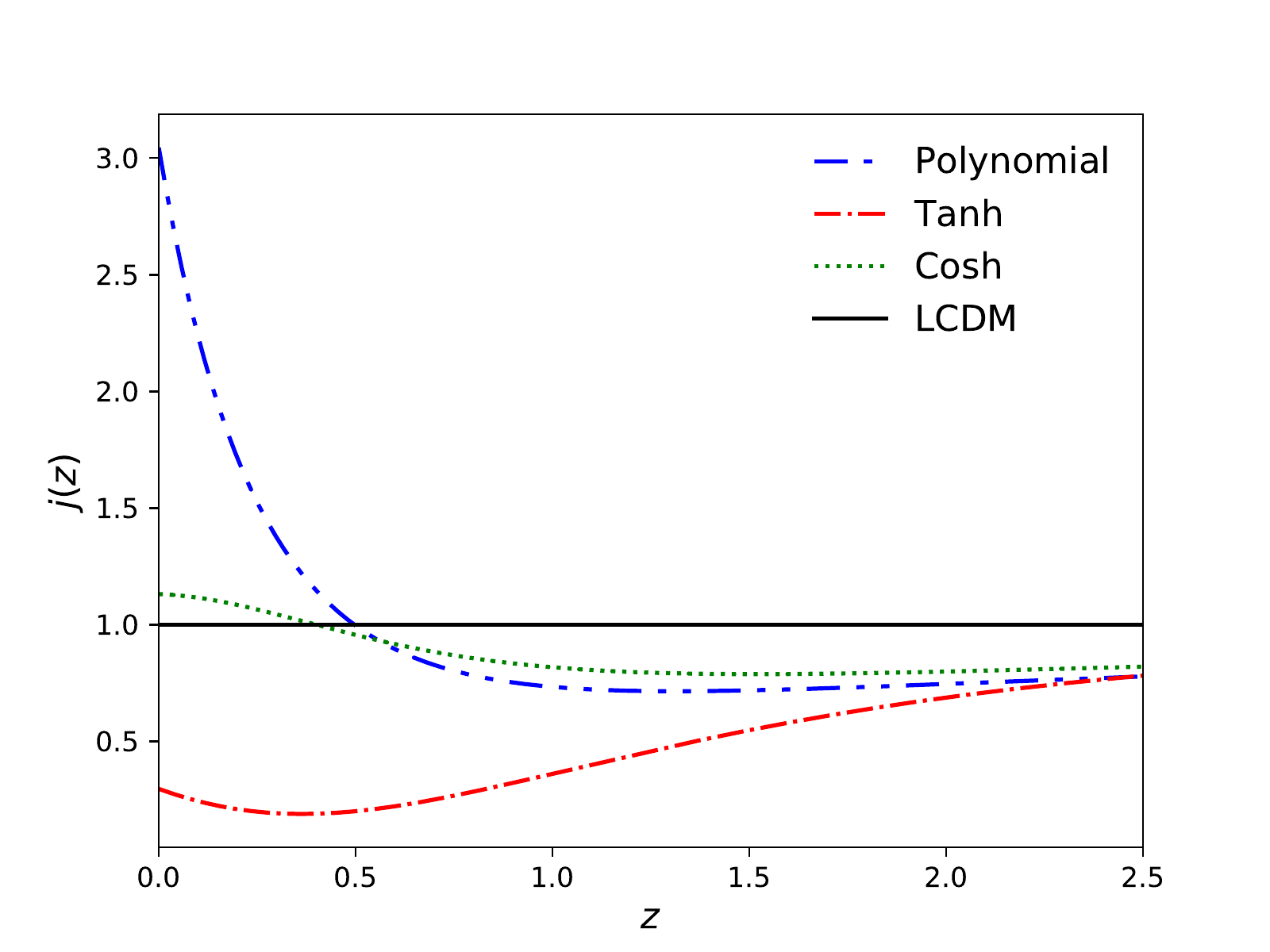}
\caption{Reconstruction of the deceleration and jerk parameter using the best fit values of the OHD$+$SNIa analysis for LCDM (black solid line), polynomial (blue double dash-dotted line), tanh (red dash-dotted line), and cosh (green dotted line) models.}
\label{fig:qzjz}
\end{figure}

Figure \ref{fig:qzjz} shows the reconstruction of the deceleration $q$ (top panel) and jerk $j$ (bottom panel) parameters with respect to the redshift in the region $0<z<2.5$.
The Universe filled by the fluids with viscosity present their deceleration-acceleration transition at earlier time ($z_t^{pol, tanh, cosh} \approx 0.75$, $ 0.82$, $0.78$) than the concordance model ($z_t^{LCDM} \approx 0.70$).
The jerk parameters of the models present the following behavior with respect to LCDM. The jerk parameter for the polynomial model has a lower-bigger transition around $z_t^{jerk}\approx 0.50$ and an increasing trend at current epochs. Similarly, the cosh model has a change from lower to bigger values around $z\approx 0.40$ but has a maximum value of less than $1.5$ at current epochs. The tanh model presents jerk values less than 1 for $z<2.5$. 

We estimate the $q$ and $j$ parameter values at current epochs 
$q_0 = -0.680^{+0.085}_{-0.102}$, $-0.539^{+0.040}_{-0.038}$, $-0.594^{+0.056}_{-0.056}$ 
and 
$j_0 = 2.782^{+1.198}_{-0.741}$, $0.297^{+0.051}_{-0.050}$, $1.124^{+0.196}_{-0.178}$ 
for the polynomial, tanh and cosh models respectively. These $q_0$ values are consistent up to $3\sigma$ CL with the one obtained for the LCDM model
($q_0^{LCDM} = -0.586^{+0.012}_{-0.011}$). With respect to the current value of the jerk, $j_0$, the polynomial (cosh) model presents a deviation  within $2.4\sigma$ ($0.7\sigma$) CL with respect to LCDM one ($j^{LCDM}=1$). In contrast, the $j_0$ of the tanh model is deviated from the LCDM one of $13.8\sigma$ CL.

The Akaike information criterion (AIC) \cite{AIC:1974, Sugiura:1978} and the Bayesian information criterion (BIC) \cite{schwarz1978}
are useful tools to compare models statistically defined by ${\rm AIC} = \chi^2 + 2k$ and ${\rm BIC} = \chi^2 + 2k\, \ln(N)$ 
respectively where $\chi^2$ is the chi-square function, $k$ is the number of estimated parameters and $N$ is the number of 
measurements. In these technique the model preferred by data is the one that has the minimum values on AIC and BIC. Hence taking 
into account the $\chi^2$ yields values reported in Tab. \ref{tab:bf_model1} and \ref{tab:bf_model23}, we obtain 
${\rm AIC}^{pol,tanh,cosh} = 1063.2$, $1063.7$, $1062.7$ and 
${\rm AIC}^{LCDM} = 1051.7$, and 
${\rm BIC}^{pol,tanh,cosh} = 1088.1$, $1083.7$, $1082.7$ and 
${\rm BIC}^{LCDM} = 1071.6$.
Following the convention presented in \cite{Liddle:2007,Davari:2018} for these criteria, we obtain that the 
data (OHD + SNIa) prefer a Universe filled with a perfect fluid (LCDM) than one filled by a viscous fluid. Between the 
phenomenological viscous models, the data prefer equally a Universe filled with a bulk viscosity modeled by a hyperbolic or a 
polynomial behaviour. Although the polynomial model is a simpler 
model than the hyperbolic one, it presents a increase behavior at the future as is shown in the statefinder phase-space.

\begin{figure}[H]
  \centering
  \includegraphics[width=.9\linewidth]{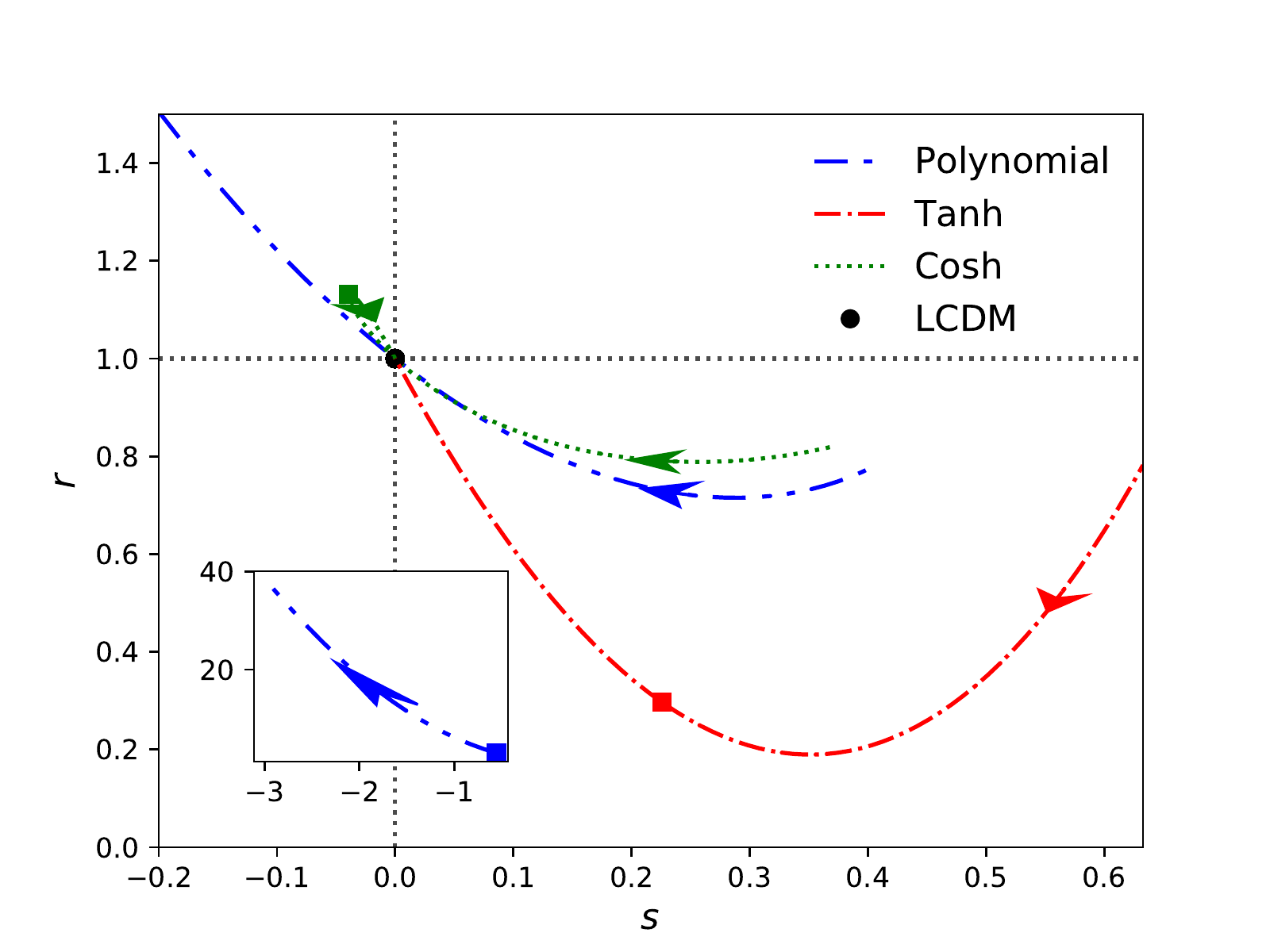}
\caption{$\{s,r\}$-statefinder diagram using the best fit values of the joint analysis in the redshift region $-1<z<2.5$. The LCDM model is represented by a black solid circle at (0,1). In blue double-dot-dashed line is the trajectory of the polynomial model, and in red dash-dotted (green dotted) line is the trajectory of the tanh (cosh) model. The blue and red (green) solid square markers over the trajectories represent the position at $z=0$. The arrows over the trajectories of the models mean the direction to the future.}
\label{fig:Statefinder}
\end{figure}

Finally, we analyze the SF diagnostic to distinguish between the behavior of the viscous models  from the LCDM model one.
In the $\{s,r\}$-SF diagram the LCDM model is a fixed point located at $\{s,r\}=(0,1)$. 
Figure \ref{fig:Statefinder} displays the $\{s,r\}$-SF space of the models in the redshift interval $-1<z<2.5$. The arrows over the trajectories of the models show the evolution direction and the solid squares are the current position ($z=0$) in the $\{s,r\}$-SF plane. We observe that the viscous models have their evolution from the quintessence region to the LCDM model. However, the polynomial model has an increasing behavior in the future with $\{s,r\}$-point $(-3,38.3)$ at $z=-1$ (see the inner plot of Fig. \ref{fig:Statefinder}), and for the tanh and cosh models end their evolution in the LCDM point.

\section{Conclusions} \label{sec:Summary}

The present work was devoted to studying the dynamics of the Universe when it is filled by a non-perfect fluid.
We analyzed three phenomenological non-perfect fluid models by introducing a bulk viscosity term in the Einstein equations through an effective pressure. 
The simplest dimensionless bulk viscosity consists of a polynomial function in terms of the redshift introduced by \cite{Xin-He:2009}.
A more complex model was proposed by \cite{FOLOMEEV200875} consisting of a bulk viscosity term of $\tanh$ function. We also studied a phenomenological alternative to the latter by modeling the bulk viscosity as a cosh form. 
Then we performed a Bayesian MCMC analysis to constrain the free parameters of the viscous models using the latest sample of OHD and SNIa and observe a good agreement of them to the data.  When they are seen in the $\{s,r\}$-statefinder space, they behave in the quintessence region. Also, the jerk parameter confirms a dynamical EoS of the viscous models. On the other hand, when we make $n=0$ for the three viscous models,  we obtain the simplest case which $\lambda = {\rm const}$ or $\xi = {\rm const}$. From our best fits, we find a deviation between the constant bulk coefficient and the tanh (cosh) model of about $4.5\sigma$ ($3.7\sigma$) and for the polynomial model, we estimate a limit of $n<-0.54$ at $99\%$ CL. When we compared statistically these models with LCDM, we find that a Universe filled by a non-perfect fluid is equally unsupported by the observational data used over the concordance model.

\section*{Acknowledgements}

The author thank the anonymous referee for thoughtful remarks and suggestions. A.H. also thanks M. Garc\'ia-Aspeitia and J. Maga\~na for useful discussions to improve the manuscript. A.H. also thank SNI Conacyt and Instituto Avanzado de Cosmolog\'ia (IAC) collaborations. The author thankfully acknowledges computer resources, technical advise and support provided by Laboratorio de Matem\'atica Aplicada y C\'omputo de Alto Rendimiento del CINVESTAV-IPN (ABACUS), Proyecto CONACYT-EDOMEX-2011-C01-165873.\\

\bibliographystyle{spphys}
\bibliography{biblio}

\end{document}